\documentclass[pdflatex,sn-mathphys,iicol]{sn-jnl}% Default with double column layout

\usepackage{csvsimple}
\usepackage{chngcntr}
\usepackage{gensymb}

\jyear{2022}%

\theoremstyle{thmstyleone}%
\theoremstyle{thmstyletwo}%

\theoremstyle{thmstylethree}%

\begin{document}

\title[Auroras on Mars]{Auroras on Mars: from Discovery to New Developments}


\author*[1]{\fnm{Dimitra} \sur{Atri}}\email{atri@nyu.edu}

\author[1]{\fnm{Dattaraj B.} \sur{Dhuri}}\email{dbd7602@nyu.edu}

\author[1]{\fnm{Mathilde} \sur{Simoni}}\email{mps565@nyu.edu}

\author[1,2]{\fnm{Katepalli R.} \sur{Sreenivasan}}\email{krs3@nyu.edu}

\affil*[1]{\orgdiv{Center for Space Science}, \orgname{New York University Abu Dhabi}, \orgaddress{\street{Saadiyat Island}, \city{Abu Dhabi}, \postcode{PO Box 129188}, \country{UAE}}}

\affil[2]{\orgdiv{Department of Physics, Courant Institute of Mathematical Sciences, and Tandon School of Engineering}, \orgname{New York University}, \orgaddress{\street{726 Broadway}, \city{New York}, \state{New York}, \country{USA}}}

%\affil[3]{\orgdiv{Department}, \orgname{Organization}, \orgaddress{\street{Street}, \city{City}, \postcode{610101}, \state{State}, \country{Country}}}

\abstract{Auroras are emissions in a planetary atmosphere caused by its interactions with the surrounding plasma environment. They have been observed in most planets and some moons of the solar system. Since their first discovery in 2005, Mars auroras have been studied extensively and is now a rapidly growing area of research. Since Mars lacks an intrinsic global magnetic field, its crustal field is distributed throughout the planet and its interactions with the surrounding plasma environment lead to a number of complex processes resulting in several types of auroras uncommon on Earth. Martian auroras have been classified as diffuse, discrete and proton aurora. With new capability of synoptic observations made possible with the Hope probe, two new types of auroras have been observed. One of them, which occurs on a much larger spatial scale, covering much of the disk, is known as discrete sinuous aurora. The other subcategory is one of proton auroras observed in patches. Further study of these phenomena will provide insights into the interactions between the atmosphere, magnetosphere and the surrounding plasma environment of Mars. We provide a brief review of the work done on the subject in the past 17 years since their discovery, and report new developments based on observations with Hope probe.}

\maketitle

\section{Introduction}\label{sec1}
Charged particles interact with a planetary atmosphere and drive a number of processes \cite{johnson2013energetic, melott2016atmospheric} involving excitation and ionization of molecules and atoms, leading to spectacular emissions in some cases, known as auroras. Auroras have been observed on several planets and some moons in the solar system. Most auroras are generated by the interaction of the solar wind or space-weather-related activity with the planetary magnetosphere (if it has one) and eventually its atmosphere. Auroras on Jovian moons are an exception because the source of charged particles are electrons accelerated by Jupiter. On Earth, auroras have been classified into three categories, each with a distinct mechanism and morphology \cite{newell2009diffuse}. The brightest and the most common type of aurora, the ``discrete aurora", is caused by the solar wind interacting with the magnetic field lines along the magnetic poles and the resulting precipitation of electrons in the atmosphere leading to these emissions. The less brighter ``diffuse aurora" occurs away from the magnetic poles caused by the scattering of particles with closed field lines in Earth's magnetosphere. These two categories are common and observed throughout the solar system. Another type is the polar rain aurora caused by solar-accelerated electrons interacting with atmosphere near the polar region. Discrete auroras are observed on planets with an intrinsic magnetic field and the diffuse ones on planets without one. Diffuse auroras are observed on Venus (even at visible wavelengths) and Jupiter's moons, Io and Europa, which lack any kind of magnetic field. Venus does not have any intrinsic magnetic field and auroras result from the interaction between the ionosphere, solar wind and other space weather-related events. They are diffuse and spread over the entire nightside of the planet \cite{phillips1986venus}. Jupiter is known to have the most powerful aurora in the solar system. They are a permanent feature observed near both magnetic poles, and also at in the region between the magnetic footprint of Io and the main auroral emission \cite{clarke1996far}. Smaller emission features are also associated with Jupiter's other moons, Europa and Ganymede. 

Due to the smaller size of Mars compared to the earth, the core of Mars cooled down early on, about 4 billion years ago and, as a result, the planet lost its intrinsic global magnetic field. The planet eventually lost most of its atmosphere, leading to drastic climate change transforming it to a cold and dry planet that we see today \cite{jakosky2018loss}. However, the magnetic field was locked in the crust at various locations on the planet, known as the crustal magnetic field \cite{langlais2008new}. These crustal fields also interact with the incoming solar wind, causing the open field lines \footnote{Although, technically, magnetic field lines always close on themselves or on a boundary, some magnetic field lines may be regarded as open or draped if one end is connected to the planet and the other end is smeared by energetic particles that enter the planetary atmosphere.} to drape the planet. Discrete auroras are caused by electrons interacting with the atmosphere along the closed field lines, whereas diffuse auroras by electrons along the open field lines. Proton auroras, caused by penetration of solar wind protons as energetic neutral atoms (ENAs), within the Martian induced magnetosphere \footnote{Since there is no global magnetic field on Mars, our use of the word loosely describes regions of influence of the magnetic field.} have also been observed \cite{deighan2018discovery}. The interaction of solar wind plasma with the Martian magnetosphere and atmosphere is very complex and leads to several types of auroral emissions which we will describe in the following sections. In addition to these emissions, charged particles also deposit energy contributing to the production of ions, photochemical changes in the atmosphere, heat, and atmospheric escape. 

In this paper we will focus on the observations of Martian auroras by various missions and their underlying mechanisms (Section \ref{sec_aurora}). In Section \ref{emm}, we will describe the new observations with the Emirates Mars Mission (EMM), also referred to as the ``Hope" probe, and discuss the importance of these observations (Section \ref{disc}). This will allow us to gain insights into a number of energetic processes in the atmosphere of Mars.

\section{Auroras on Mars}\label{sec_aurora}
\subsection{Discrete auroras}
Discrete auroras are the most widely studied auroras on Mars. They are confined to small spatial scales and generally observed near areas with strong crustal magnetic fields. They were first discovered in 2005 \cite{bertaux2005discovery} using the SPICAM (Spectroscopy for the Investigation of the Characteristics of the Atmosphere of Mars) \cite{bertaux2006spicam} instrument, a UV spectrometer on board Mars Express \cite{chicarro2004mars}. The enhanced emission was observed across several bands -- the CO Band (135-170 nm), CO Cameron bands (190-270 nm), CO$_{2}$+ UV Doublet UVD (289 nm), and oxygen line at 297.2 nm, generated by electrons interacting with CO$_{2}$. 

These auroras are thought to result from electrons interacting with the upper atmosphere at locations with magnetic field anomalies in the crust of Mars. A large flux of electrons move along the crustal field lines and excite the upper atmosphere, leading to these emissions. This increased concentration of electrons is highly localized, as opposed to auroras observed on Earth, hence the localized nature of the emissions. They are observed at an altitude of around 130 km in patches of around 10 km. The main mechanism is the precipitation of electrons of energies from 40-200 eV in the night-side atmosphere at an altitude of around 135 km. The peak energy distribution of these electrons is similar to that of electrons responsible for discrete auroras on Earth \cite{brain2006origin}. Observations made by the Mars Global Surveyor (MGS) orbiter led to the spectra of these electrons and reported peak energies in the 100 eV - 2.5 keV range \cite{brain2006origin}. The flux was enhanced by a factor of 10-10,000 at the peak compared to the night-side electron spectra. 

Discrete auroras have also been studied extensively with the MAVEN (Mars Atmosphere and Volatile and EvolutioN \cite{jakosky2015mars}) orbiter with the IUVS (Imaging Ultraviolet Spectrograph) instrument \cite{schneider2021discrete}. These auroras are common and known to occur every evening. The MAVEN observations reported events of high intensity and occurrence rates in the southern hemisphere, populated by strong crustal fields. These observations also included events in regions with low or no crustal fields, although the auroras were less bright---though they had the same spectral properties and peak altitude. A strong time dependence was noted, the main occurrence taking place during evening hours when the Interplanetary Magnetic Field (IMF) is favorably aligned. Relatively few observations have been made in the northern hemisphere where the crustal fields are known to be weak. 

Empirical study \cite{xu2022empirically} of auroral electron events, which are generated by electrons in the 50 eV - 2 keV range and a fluence of above 1.8E-5 W/m2/sr, found a linear scaling between electron flux and brightness of aurora events. The recent paper \cite{girazian2022discrete} studied the dependence of the frequency of aurora occurrence on upstream solar wind conditions --- in particular, on the IMF strength and cone angle, and solar wind dynamic pressure. The authors found that there is a distinct difference between the dependence of these quantities on the occurrence rate within and outside the Strong Crustal Field Region (SCFR) located in the southern hemisphere of Mars. A high occurrence rate was reported in SCFR for negative IMF clock angles, increasing the likelihood of magnetic reconnection. This leads to a larger number of particles entering the lower atmosphere and producing auroras. It was also found that the IMF strength increases the occurrence rate moderately, whereas the IMF orientation has no impact on the detection frequency outside of the SCFR. The frequency strongly depends on the solar wind dynamic pressure although it has very little impact on its brightness outside the region. The occurrence rate moderately depends on the IMF clock and cone angle, and the solar wind dynamic pressure within the SCFR.

\subsection{Diffuse auroras}
Diffuse auroras \cite{schneider2015discovery} were first discovered by the IUVS (Imaging Ultraviolet Spectrograph) instrument \cite{mcclintock2015imaging} on board the MAVEN \cite{jakosky2015mars} spacecraft. Emissions were observed at low altitudes, coincident with Solar Energetic Particle (SEP) events, down to 60 km altitude, indicating that the source consists of higher energy particles than those producing discrete aurora at higher altitudes. As opposed to discrete aurora, where particles are accelerated by interactions with the magnetosphere, diffuse aurora are produced by electrons accelerated at the Sun, not locally. These solar accelerated electrons strike the open and draped field lines on Mars, interact with the atmosphere, and produce the emissions. 

The intensity of diffuse auroras is about 2 orders of magnitude lower than typical dayglow. They are seen at about 60 km altitude, as opposed to dayglow, which occurs at altitudes between 120 to 150 km. Further analysis has shown that the energy of these particles is 2-3 orders of magnitude higher than those producing discrete auroras, and so they penetrate much deeper into the atmosphere. There is no correlation between the geographic location of the planet and the observed brightness. This is because the location of the open and draped magnetic field around the planet changes as it encounters variable solar wind conditions. Several occurrences of auroras have been reported during SEP events. The averaged vertical emission profile shows a peak at 60-70 km altitude. Modeling suggests that they are caused by electrons in the 10-200 keV range with a power law index of -2.2.

Polar rain aurora seen on the Earth are also produced by solar-accelerated electrons just as in case of Mars. Venus and Jupiter's moons Io and Europa also see similar diffuse auroras distributed throughout the planet. Venus does not possess any magnetic field, either global or discrete and auroras are caused by SEPs and are visible in the UV and visible wavelengths. In case of Europa and Io, electrons accelerated by the Jovian magnetic field interact with the thin atmosphere of the moons and produce diffuse auroras distributed over the globe.

\subsection{Proton auroras}
In addition to electrons, high-energy solar wind protons also interact with the atmospheres of planets and lead to auroral emissions. Earth's proton aurora observations are known for almost a century with emissions in H Balmer-$\rm{\alpha}$ and -$\rm{\beta}$ lines. However, on Mars, they have been first identified only recently in 2017 from IUVS Lyman-$\rm{\alpha}$ observations \cite{deighan2018discovery}. Proton auroras have been identified subsequently also from SPICAM \cite{ritter2018proton} and very recently from EMM/EMUS \cite{holsclaw2021emirates,chaffin2022patchy} with novel morphological features (see Section \ref{emm}). The lack of a global intrinsic magnetic field makes proton auroras on Mars very different from those on Earth.

As a consequence of weak gravity and the absence of a strong planetary magnetic field, H corona on Mars extends well outside the induced magnetosphere and is exposed to the solar wind. Solar wind protons can undergo charge exchange with the neutral H in corona, outside the bow shock, and become Energetic Neutral Atoms (ENAs). These ENAs can penetrate the bow shock and induced magnetosphere to reach thermospheric altitudes (100-200~km). Inside the thermosphere, ENAs interact with the $\rm{CO_2}$ atmosphere to convert back-and-forth between protons and ENAs through electron-stripping and charge exchange, respectively. These high energy ($\sim$ 1~keV) solar wind penetrating protons have been observed with MAVEN \cite{halekas2015wind}. The penetrating protons, which get excited in this process, deexcite via the release of the proton aurora emissions \cite{deighan2018discovery}. The characteristic signature of proton auroras on Mars is the several kR intensity enhancement in Ly-$\rm{\alpha}$, first clearly seen in IUVS limb scan altitude profiles as shown in Panel (b) of Figure~\ref{fig:auroras_proton}. With EMUS these proton aurora are identified as prominent intensity enhancement in Ly-$\rm{\alpha}$ and Ly-$\rm{\beta}$ observations as shown in Panel (a) of Figure~\ref{fig:auroras_proton}. Unlike discrete and diffuse electron auroras, which occur on the nightside, proton auroras are mostly observed on the dayside \cite{hughes2019proton}. Proton aurora emissions are typically several times more intense than the characteristic background H Ly-$\rm{\alpha}$ dayglow (as shown in Figure~\ref{fig:auroras_proton}).

\begin{figure*}
\centering
	\includegraphics[width=\textwidth,trim={2cm 2cm 2cm 2cm},clip]{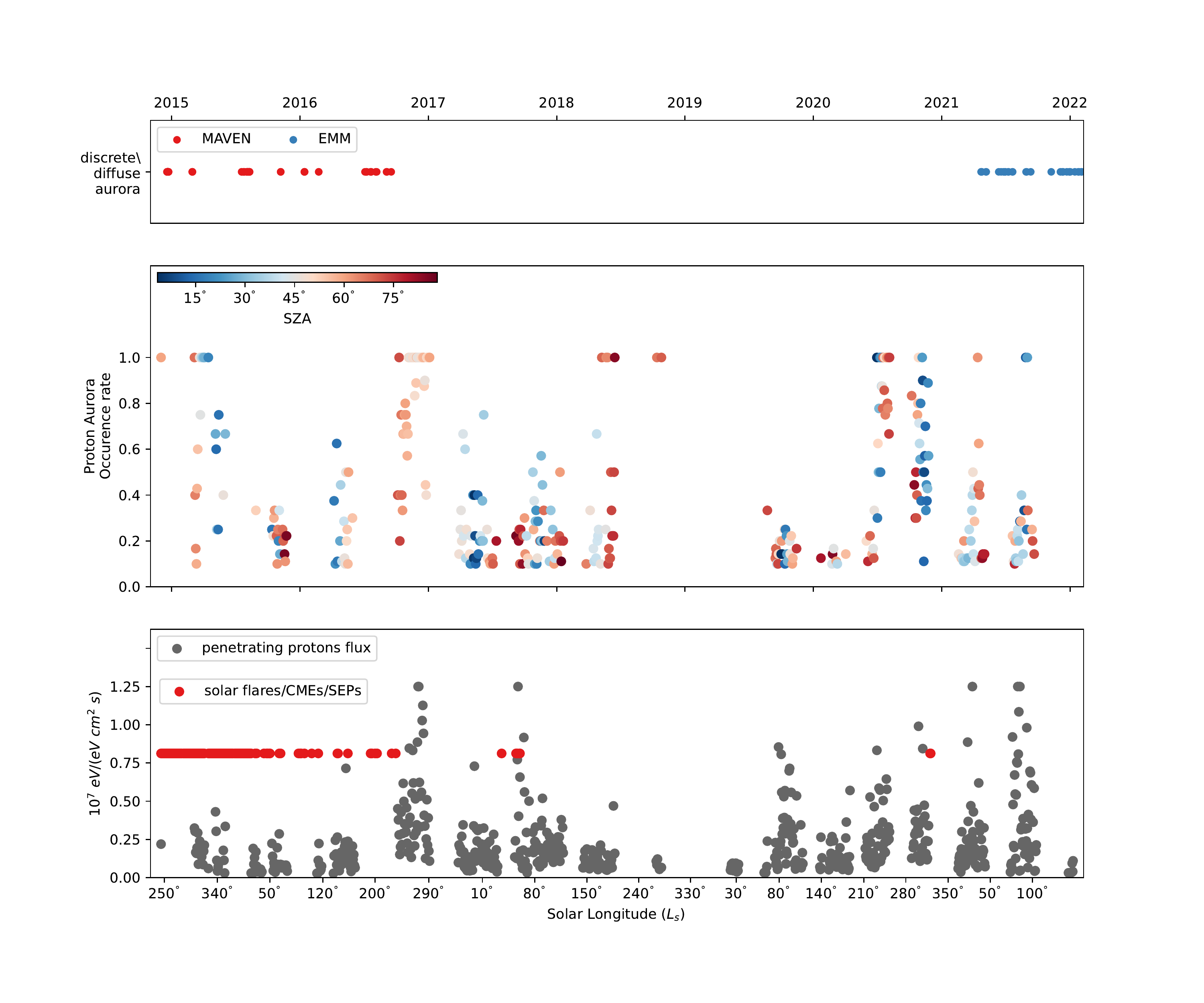}
    \caption{Factors influencing Mars proton aurora activity. Occurrence rates of proton auroras observed by MAVEN/IUVS and identified according to the criterion by Hughes et al. \cite{hughes2019proton} are shown as a function of solar season on Mars. Proton aurora occur relatively frequently with occurrence rates of almost unity close to $\rm{L_s=270}$. Flux of penetrating solar wind protons, identified in MAVEN/SWIA observations, following the algorithm from Halekas et al. \cite{Halekas2017variability}, is also shown. The peak proton aurora occurrence rates coincide with an increased flux of penetrating protons, due to an inflated H corona around southern summer solstice. Solar activity events observed by MAVEN, as well as the discrete and diffuse electron aurora events observed by MAVEN and EMM, are also shown.
    Frequency of electron aurora observations is significantly increased with the global coverage of EMM/EMUS compared to MAVEN.}
    \label{fig:auroras_factors}
\end{figure*}

Proton auroras are one of the most widely observed auroras on Mars, found in $14\%$ of periapsis limb scan observations from IUVS \cite{hughes2019proton}. Since the ENAs responsible for proton auroras are formed in H corona, the occurrence rates of proton aurora peak to near unity around southern summer solstice $\rm{L_s=270\degree}$, when the H corona is inflated with increased coronal H column densities. Figure~\ref{fig:auroras_factors} shows proton aurora (identified in IUVS periapsis limb scans) occurrence rate  variation as a function season with a distinct peak at $\rm{L_s=270\degree}$. This peak corresponds to an increased flux of penetrating solar wind protons (as observed by MAVEN, also shown in Figure~\ref{fig:auroras_factors}) following the inflated H corona and proximity to perihelion \cite{Halekas2017variability}. Additionally, solar activity (flares/CMEs) tend to increase the intensity of proton aurora emissions. While strong (crustal) magnetic fields are expected to impede penetrating protons \cite{GERARD2019mag}, proton auroras possibly occur more frequently during radial orientation of IMF \cite{Hughes2021AGU}. Full implications of IMF orientation and Mars crustal magnetic fields on proton aurora occurrences and morphology remain to be studied (see Section \ref{patchy}).

\section{Recent observations with the Hope probe}\label{emm}
While Martian auroras have been studied since 2005, observations were confined to localized patches on the planet due to the limited geographical coverage of orbiters such as MEX and MAVEN. On the other hand, because of its high-altitude (19,970 km $\times$ 42,650 km) orbit ($\sim$55 hr period), EMM \cite{amiri2022emirates, almatroushi2021emirates} provides a global view of the planet with a high sensitivity in the visible \cite{2022AoM}, infrared \cite{atri2022diurnal}, and UV wavelengths \cite{holsclaw2021emirates}. This new capability has enabled the observation of auroras on a planetary scale, and has lead to the discovery of two new types of auroras, which we describe below.

\subsection{Sinuous auroras}
EMUS \cite{holsclaw2021emirates} is a UV spectrometer on board EMM, sensitive in the 100-170 nm wavelength range. The orbiter has been operational around Mars since February 2021, and EMUS observations used here are from the science phase of the mission starting from May 2021. EMUS has enabled the first observations of auroras in the extreme UV (EUV: 10-120 nm) and far UV (FUV: 122-200 nm), which can be observed in about three-fourths of all nightside observations \cite{lillisfirst}. Discrete auroras can be observed best in the 130.4 nm oxygen line, which results from the decay of excited oxygen from 3S1 to 3P2. These excited oxygen atoms are produced both by direct interaction with energetic electrons and by dissociative excitation of CO$_{2}$ and CO. They have appeared in 31 EMM observations identified in Table \ref{table1}.

\begin{figure*}
\centering
	\includegraphics[width=12cm]{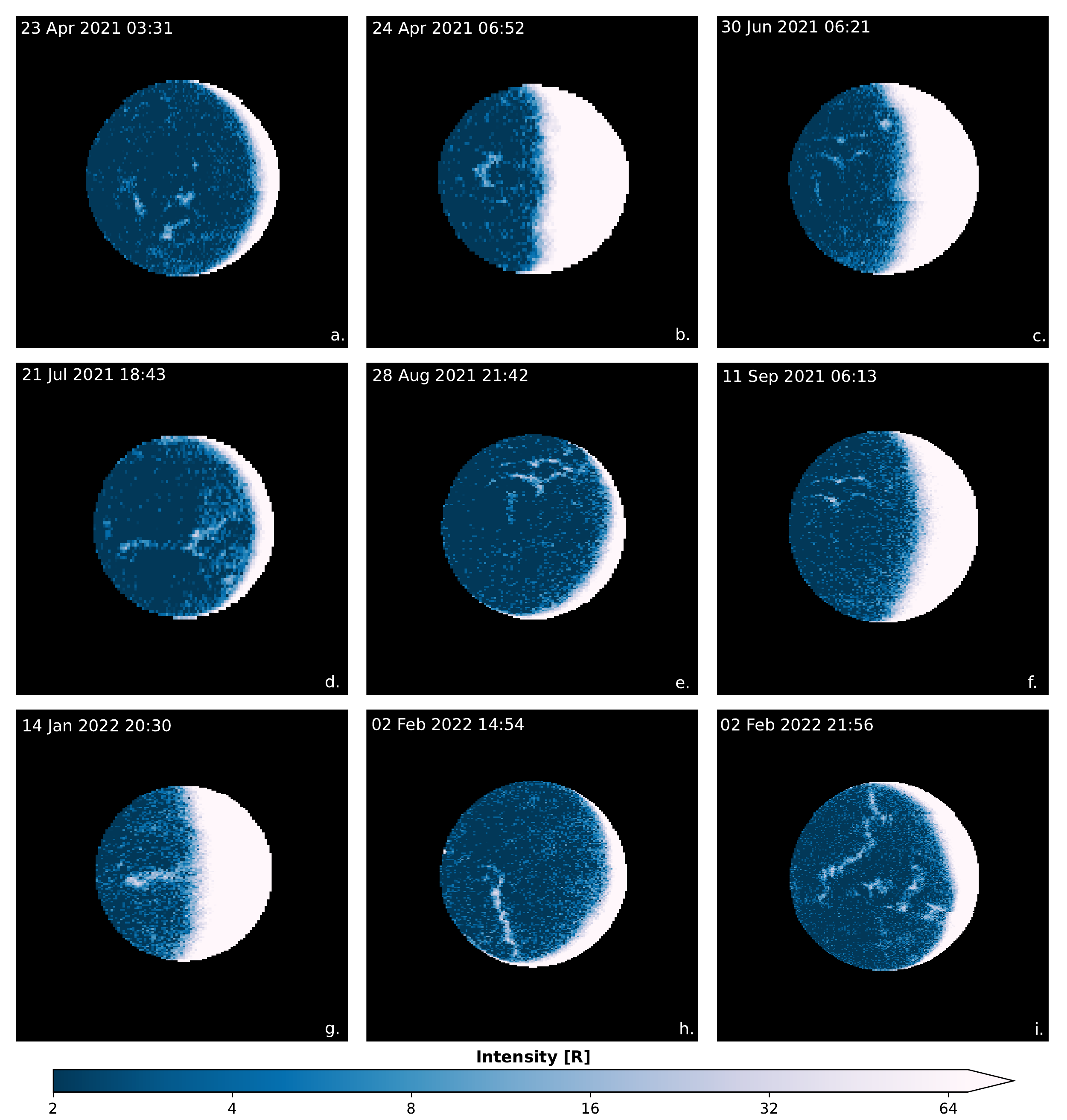}
    \caption{Selected observations of discrete auroras produced with data from EMM/EMUS in the 130.4 nm oxygen band. The data of each swath was first extracted from the latest versions of L2B data level files with OS2 observation mode \cite{holsclaw2021emirates}. The log$_{2}$ function was subsequently applied to the images. Sinuous discrete auroras can be seen in images d, g, h and i. Additional observations are available in the Appendix, Figures \ref{fig:auroras_observations_appendix1}, \ref{fig:auroras_observations_appendix2}, and \ref{fig:auroras_observations_appendix3}.}
    \label{fig:auroras_observations}
\end{figure*}

Figure \ref{fig:auroras_observations} shows 9 selected observations from the EMUS instrument in the 130.4 nm oxygen band. They display discrete auroral features captured during the orbits 42, 43, 73, 83, 99, 105, 160, 167, and 168, from April 2021 to February 2022.
%The raw data was first extracted from the latest versions of L2b data level files with os2 calibration mode. The log$_{2}$ function was subsequently applied to the images. 
The auroras can be observed with their lighter color on the night side of the Mars disk.  Additionally, 22 other observations with discrete auroral events can be found in Figures \ref{fig:auroras_observations_appendix1}, \ref{fig:auroras_observations_appendix2}, and \ref{fig:auroras_observations_appendix3} of the Appendix.

\begin{figure*}
\centering
	\includegraphics[width=12cm]{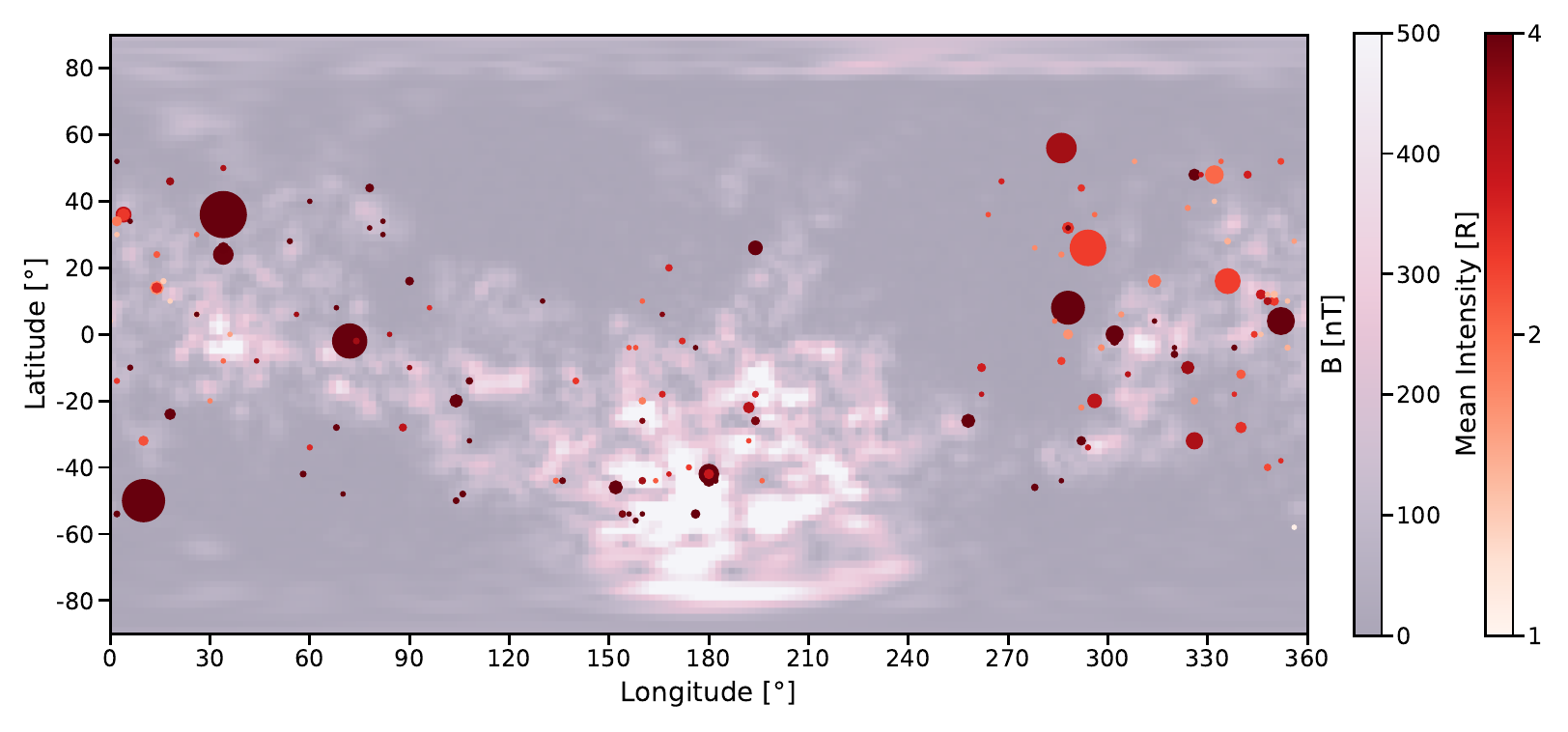}
    \caption{Crustal magnetic field at 120 km altitude and discrete auroras observed by EMM/EMUS. The data were extracted from 31 observations with visible discrete auroras (identified in Table \ref{table1}), between April 2021 and February 2022. Each auroral event is represented by a circle, the center of which is located at the centroid of the aurora and the radius of which is proportional to the aurora size. In addition, the color of each circle indicates the aurora mean intensity. The Mars crustal magnetic field is from Gao et al. \cite{Gao2021crustal}.}
    \label{fig:auroras_magnetic_field}
\end{figure*}

Auroral features can also be observed in other wavelengths, such as 98.9 nm, 102.7 nm, 104 nm and 135.6 nm, but the signals are weaker and noisy. The auroras of 130.4 nm line spreads in an elongated serpentine structure over thousands of kilometers, covering much of the disk. As can be seen in Figure \ref{fig:auroras_magnetic_field} their occurrence rate is the highest in regions where the crustal field is weak and/or vertical. The crustal field data used in the figure was obtained from a recent study which combines MAVEN and MGS datasets \cite{Gao2021crustal}.  However, the brightest ones are seen in regions with the strongest crustal fields, as was the case with IUVS observations as well. Strong vertical crustal field lines facilitate the interaction of electrons with the atmosphere. 
%In these regions, field lines are open, i.e. connect the solar wind and magnetotail to the atmosphere, where energetic solar wind electrons precipitate leading to auroral emissions.
Overall, these observations can be grouped into three categories, crustal field aurora, non-crustal field patchy discrete aurora, and sinuous discrete aurora \cite{lillisfirst}. Although the origin of the last category of auroras is unknown, it has been speculated that they originate from the magnetotail current sheet, which could be accelerating electrons to the nightside atmosphere.

\begin{table}[!ht]
    \centering
    \begin{tabular}{|l|l|}
    \hline
        \textbf{Orbit} & \textbf{Time} \\ \hline
        42 & 22 Apr 2021 23:40 \\ \hline
        42 & 23 Apr 2021 03:31 \\ \hline
        43 & 24 Apr 2021 06:52 \\ \hline
        48 & 06 May 2021 19:24 \\ \hline
        48 & 07 May 2021 12:37 \\ \hline
        64 & 12 Jun 2021 02:45 \\ \hline
        68 & 18 Jun 2021 23:12 \\ \hline
        71 & 25 Jun 2021 22:51 \\ \hline
        73 & 30 Jun 2021 06:21 \\ \hline
        73 & 30 Jun 2021 12:19 \\ \hline
        77 & 09 Jul 2021 11:25 \\ \hline
        77 & 09 Jul 2021 15:21 \\ \hline
        82 & 21 Jul 2021 01:11 \\ \hline
        83 & 21 Jul 2021 18:43 \\ \hline
        99 & 28 Aug 2021 21:42 \\ \hline
        105 & 11 Sep 2021 06:13 \\ \hline
        131 & 08 Nov 2021 13:14 \\ \hline
        142 & 04 Dec 2021 17:27 \\ \hline
        145 & 10 Dec 2021 08:53 \\ \hline
        145 & 12 Dec 2021 05:13 \\ \hline
        150 & 22 Dec 2021 23:10 \\ \hline
        154 & 30 Dec 2021 16:06 \\ \hline
        154 & 30 Dec 2021 21:12 \\ \hline
        154 & 01 Jan 2022 17:32 \\ \hline
        160 & 14 Jan 2022 20:30 \\ \hline
        164 & 24 Jan 2022 12:18 \\ \hline
        165 & 26 Jan 2022 05:59 \\ \hline
        167 & 02 Feb 2022 14:54 \\ \hline
        168 & 02 Feb 2022 21:56 \\ \hline
        177 & 25 Feb 2022 01:03 \\ \hline
        177 & 25 Feb 2022 09:44 \\ \hline 
    \end{tabular}
    \vspace{0.5cm}
    \caption{Orbit and times of 31 auroral events observed by EMM/EMUS in the 130.4 nm oxygen band.}
    \label{table1}
\end{table}

\begin{figure*}
\centering
	\includegraphics[width=\textwidth,trim={7cm 1.4cm 4.5cm 3cm},clip]{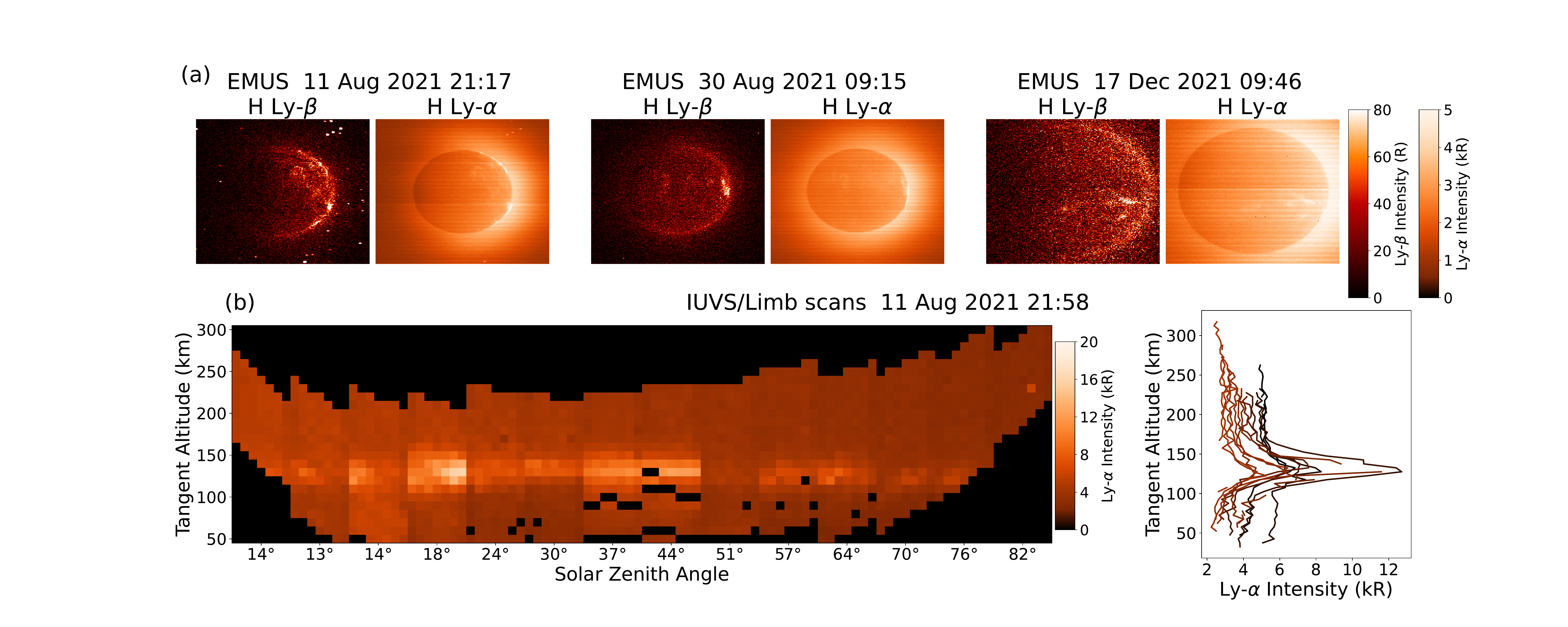}
    \caption{Example proton aurora observations from EMM/EMUS (panel a) and MAVEN/IUVS (panel b). Proton auroras are seen as (dayside) enhancements in Ly-$\rm{\alpha}$ (121.6 nm) from both EMM/EMUS and MAVEN/IUVS observations as well as Ly-$\rm{\beta}$ (102.6 nm) observations from EMM/EMUS. The MAVEN/IUVS alitude profile (panel b right) shows a clear enhancement peak at altitudes between 110-150~km. The EMM/EMUS data shown here is from OS2 and OS1 observation modes \cite{holsclaw2021emirates} and MAVEN/IUVS observations are from the periapsis limb scans. Proton auroras occur relatively frequently on Mars and are widely seen in MAVEN/IUVS observations (see Figure~\ref{fig:auroras_factors}).}  
    \label{fig:auroras_proton}
\end{figure*}

\subsection{Patchy proton auroras}\label{patchy}

As discussed earlier, the proton auroras on Mars are typically produced by solar wind protons penetrating the magnetosphere as ENAs and precipitating at the thermosphere. The penetrating proton flux is primarily determined by solar wind energy and density as well as the column density of H corona. All of these (on dayside) typically are relatively uniform over Mars and therefore proton aurora emissions are generally expected to be uniform on the dayside disk. Previously, variability in the proton aurora brightness (and altitude profiles) has been observed in IUVS detections, obtained across different limb scan observations taken during the same periapsis pass \cite{deighan2018discovery,hughes2019proton}. However, this variability is indistinguishable from the temporal variability as MAVEN travels significantly in its orbit while capturing different limb scan observations. The proton auroras detected by EMUS (shown in panel a Figure~\ref{fig:auroras_proton}) for the first time present a definite evidence of localization of proton aurora emissions (hence called patchy proton auroras) over the dayside \cite{chaffin2022patchy}. Although the patchy proton auroras are less frequently observed with EMUS (about 2\% of observations), compared to IUVS observation rate of 14\% \cite{hughes2019proton}, they reveal significantly varied interactions of solar wind with the induced magnetosphere on Mars. 

The observed spatial patchiness in auroral emissions may be caused by the disorganization of the magnetosphere when IMF is occasionally radial with the magnetic field, whose direction is parallel to the flow \cite{crismani2019ionosphere,chaffin2022patchy}. This leads to localized solar wind proton precipitation causing patchy auroral emissions. During the EMUS patchy proton aurora observation on 11 August 2021, the MAVEN observation at approximately the same time showed weaker and highly fluctuating magnetic field in the sheath, suggesting a possible radial orientation of the IMF upstream of the bow shock \cite{chaffin2022patchy}. MAVEN observations also show solar wind energy protons (keV) near thermospheric altitudes as a direct evidence for the precipitated protons, confirming the patchy proton aurora. Apart from radial IMF orientations, EMUS patchy auroras have also been observed as a result of localized plasma turbulence from a locally perpendicular IMF orientation with respect to the bow shock \cite{chaffin2022patchy}. The proton precipitation and aurora are therefore limited to only certain locations on the dayside where these conditions are met (e.g. 30 August 2021 observation in Figure~\ref{fig:auroras_proton}). The patchy auroras, formed during radial IMF conditions, have also shown a correlation with the crustal magnetic fields, with the auroral emissions observed mainly in regions of weaker crustal fields, suggesting that the solar wind plasma precipitates easily at the weak crustal field regions in contrast to strong crustal field regions that can block the proton precipitation \cite{GERARD2019mag}. Combined EMUS and MAVEN observations in near future will shed further light on the phenomena of the patchy proton auroras, the role of IMF and crustal magnetic field orientations in localized solar wind proton precipitation, and on their consequences for the variability in solar wind and Mars magnetosphere interactions.

\section{Discussion}\label{disc}
Auroras are a result of the interplay between solar wind, planetary magnetosphere and atmosphere, and are observed on most planets of the Solar system. Jupiter has a very strong magnetic field and auroral emissions are a permanent feature of the planet at the magnetic poles. Venus, on the other hand, does not have a magnetic field; yet, diffuse auroras occur on the planet during major space weather events. Mars falls between the two categories of magnetized (Earth, Jupiter) and unmagnetized (Venus) planets. Since their discovery in 2005, our understanding of Martian auroras has increased considerably, especially with new observations by EMM but have also given rise to a number of questions about the complex interaction between solar wind plasma with the Martian magnetosphere and atmosphere.  

While regular EUV photons are responsible for the background airglow emission seen on Mars, discrete auroras are caused by electron interactions, carrying up to 1 keV energy. Diffuse auroras, on the other hand, are driven by electrons of much higher energy, between 10 to 200 keV. Discrete auroras are caused by particle acceleration locally, whereas diffuse auroras due to energetic particles are accelerated by the Sun. Magnetic field configuration also plays an important part in auroral emissions. When the magnetic field lines are closed, both the ends are connected to the planet, the solar wind particles undergo acceleration along these field lines and cause discrete auroras. When the magnetic field lines are open or draped, the energetic particles can access the planetary atmosphere through them, causing diffuse auroras. Further studies of the newly discovered patchy proton aurora phenomena with EMM, caused by localized precipitations of solar wind protons inside the magnetosphere, promise to uncover new physical processes in the solar wind interaction with the Martian atmosphere.

While MAVEN is equipped with instruments that carry out {\it in situ} measurements of charged particles interacting with Mars, and measure the resulting emissions, its geographic and temporal coverage is quite limited. EMM provides a synoptic view of the planet with high sensitivity in the UV, but lacks the capability of {\it in situ} measurements. A combined analysis of data from these two missions will allow a more comprehensive study of auroras. Initial results from EMM, which we have reported here, have revealed new subcategories of auroras. More work needs to be done in order to identify the underlying physics. As we have described earlier, the original discovery of auroras was made by MEX which has been observing Mars from the orbit since December 2003 till date. Its capabilities to study these new auroras, especially alongside EMM and MAVEN, need to be explored further.

The auroral emissions comprise only a small fraction of the overall energy of the incident particles. These complex electromagnetic interactions also lead to photochemical changes in the atmosphere and deposit heat, contributing to atmospheric escape. This aspect also needs to be investigated further in order to understand how Mars lost most of its atmosphere, leading to drastic climate change. More research on auroras will enable us to better understand these processes occurring at present, and extrapolate them back in time to better comprehend the evolution of the planet.

\section*{Acknowledgments}
This work was supported by the New York University Abu Dhabi (NYUAD) Institute Research Grant G1502 and the ASPIRE Award for Research Excellence (AARE) Grant S1560 by the Advanced Technology Research Council (ATRC). EMM data was obtained from the EMM Science Data Center (SDC) and MAVEN data from the Planetary Data System (PDS). Data analysis was performed on the NYUAD High Performance Computing (HPC) resources. It is a pleasure to contribute this article to the special issue in honor of Professor Kurt Becker.

\section*{Statements and Declarations}
The authors have no competing interests to declare that are relevant to the content of this article.
\bibliography{aurora_mars}% common bib file


\begin{thebibliography}{33}
\ifx \bisbn   \undefined \def \bisbn  #1{ISBN #1}\fi
\ifx \binits  \undefined \def \binits#1{#1}\fi
\ifx \bauthor  \undefined \def \bauthor#1{#1}\fi
\ifx \batitle  \undefined \def \batitle#1{#1}\fi
\ifx \bjtitle  \undefined \def \bjtitle#1{#1}\fi
\ifx \bvolume  \undefined \def \bvolume#1{\textbf{#1}}\fi
\ifx \byear  \undefined \def \byear#1{#1}\fi
\ifx \bissue  \undefined \def \bissue#1{#1}\fi
\ifx \bfpage  \undefined \def \bfpage#1{#1}\fi
\ifx \blpage  \undefined \def \blpage #1{#1}\fi
\ifx \burl  \undefined \def \burl#1{\textsf{#1}}\fi
\ifx \doiurl  \undefined \def \doiurl#1{\url{https://doi.org/#1}}\fi
\ifx \betal  \undefined \def \betal{\textit{et al.}}\fi
\ifx \binstitute  \undefined \def \binstitute#1{#1}\fi
\ifx \binstitutionaled  \undefined \def \binstitutionaled#1{#1}\fi
\ifx \bctitle  \undefined \def \bctitle#1{#1}\fi
\ifx \beditor  \undefined \def \beditor#1{#1}\fi
\ifx \bpublisher  \undefined \def \bpublisher#1{#1}\fi
\ifx \bbtitle  \undefined \def \bbtitle#1{#1}\fi
\ifx \bedition  \undefined \def \bedition#1{#1}\fi
\ifx \bseriesno  \undefined \def \bseriesno#1{#1}\fi
\ifx \blocation  \undefined \def \blocation#1{#1}\fi
\ifx \bsertitle  \undefined \def \bsertitle#1{#1}\fi
\ifx \bsnm \undefined \def \bsnm#1{#1}\fi
\ifx \bsuffix \undefined \def \bsuffix#1{#1}\fi
\ifx \bparticle \undefined \def \bparticle#1{#1}\fi
\ifx \barticle \undefined \def \barticle#1{#1}\fi
\bibcommenthead
\ifx \bconfdate \undefined \def \bconfdate #1{#1}\fi
\ifx \botherref \undefined \def \botherref #1{#1}\fi
\ifx \url \undefined \def \url#1{\textsf{#1}}\fi
\ifx \bchapter \undefined \def \bchapter#1{#1}\fi
\ifx \bbook \undefined \def \bbook#1{#1}\fi
\ifx \bcomment \undefined \def \bcomment#1{#1}\fi
\ifx \oauthor \undefined \def \oauthor#1{#1}\fi
\ifx \citeauthoryear \undefined \def \citeauthoryear#1{#1}\fi
\ifx \endbibitem  \undefined \def \endbibitem {}\fi
\ifx \bconflocation  \undefined \def \bconflocation#1{#1}\fi
\ifx \arxivurl  \undefined \def \arxivurl#1{\textsf{#1}}\fi
\csname PreBibitemsHook\endcsname

%%% 1
\bibitem{johnson2013energetic}
\begin{bbook}
\bauthor{\bsnm{Johnson}, \binits{R.E.}}:
\bbtitle{Energetic Charged-particle Interactions with Atmospheres and Surfaces}
vol. \bseriesno{19}.
\bpublisher{Springer},
\blocation{Berlin}
(\byear{2013})
\end{bbook}
\endbibitem

%%% 2
\bibitem{melott2016atmospheric}
\begin{barticle}
\bauthor{\bsnm{Melott}, \binits{A.L.}},
\bauthor{\bsnm{Thomas}, \binits{B.C.}},
\bauthor{\bsnm{Laird}, \binits{C.M.}},
\bauthor{\bsnm{Neuenswander}, \binits{B.}},
\bauthor{\bsnm{Atri}, \binits{D.}}:
\batitle{Atmospheric ionization by high-fluence, hard-spectrum solar proton
  events and their probable appearance in the ice core archive}.
\bjtitle{Journal of Geophysical Research: Atmospheres}
\bvolume{121}(\bissue{6}),
\bfpage{3017}--\blpage{3033}
(\byear{2016})
\end{barticle}
\endbibitem

%%% 3
\bibitem{newell2009diffuse}
\begin{botherref}
\oauthor{\bsnm{Newell}, \binits{P.}},
\oauthor{\bsnm{Sotirelis}, \binits{T.}},
\oauthor{\bsnm{Wing}, \binits{S.}}:
Diffuse, monoenergetic, and broadband aurora: The global precipitation budget.
Journal of Geophysical Research: Space Physics
\textbf{114}(A9)
(2009)
\end{botherref}
\endbibitem

%%% 4
\bibitem{phillips1986venus}
\begin{barticle}
\bauthor{\bsnm{Phillips}, \binits{J.}},
\bauthor{\bsnm{Stewart}, \binits{A.}},
\bauthor{\bsnm{Luhmann}, \binits{J.}}:
\batitle{The venus ultraviolet aurora: Observations at 130.4 nm}.
\bjtitle{Geophysical Research Letters}
\bvolume{13}(\bissue{10}),
\bfpage{1047}--\blpage{1050}
(\byear{1986})
\end{barticle}
\endbibitem

%%% 5
\bibitem{clarke1996far}
\begin{barticle}
\bauthor{\bsnm{Clarke}, \binits{J.T.}},
\bauthor{\bsnm{Ballester}, \binits{G.E.}},
\bauthor{\bsnm{Trauger}, \binits{J.}},
\bauthor{\bsnm{Evans}, \binits{R.}},
\bauthor{\bsnm{Connerney}, \binits{J.E.}},
\bauthor{\bsnm{Stapelfeldt}, \binits{K.}},
\bauthor{\bsnm{Crisp}, \binits{D.}},
\bauthor{\bsnm{Feldman}, \binits{P.D.}},
\bauthor{\bsnm{Burrows}, \binits{C.J.}},
\bauthor{\bsnm{Casertano}, \binits{S.}}, \betal:
\batitle{Far-ultraviolet imaging of jupiter's aurora and the io
  “footprint”}.
\bjtitle{Science}
\bvolume{274}(\bissue{5286}),
\bfpage{404}--\blpage{409}
(\byear{1996})
\end{barticle}
\endbibitem

%%% 6
\bibitem{jakosky2018loss}
\begin{barticle}
\bauthor{\bsnm{Jakosky}, \binits{B.}},
\bauthor{\bsnm{Brain}, \binits{D.}},
\bauthor{\bsnm{Chaffin}, \binits{M.}},
\bauthor{\bsnm{Curry}, \binits{S.}},
\bauthor{\bsnm{Deighan}, \binits{J.}},
\bauthor{\bsnm{Grebowsky}, \binits{J.}},
\bauthor{\bsnm{Halekas}, \binits{J.}},
\bauthor{\bsnm{Leblanc}, \binits{F.}},
\bauthor{\bsnm{Lillis}, \binits{R.}},
\bauthor{\bsnm{Luhmann}, \binits{J.}}, \betal:
\batitle{Loss of the martian atmosphere to space: Present-day loss rates
  determined from maven observations and integrated loss through time}.
\bjtitle{Icarus}
\bvolume{315},
\bfpage{146}--\blpage{157}
(\byear{2018})
\end{barticle}
\endbibitem

%%% 7
\bibitem{langlais2008new}
\begin{barticle}
\bauthor{\bsnm{Langlais}, \binits{B.}},
\bauthor{\bsnm{Quesnel}, \binits{Y.}}:
\batitle{New perspectives on mars’ crustal magnetic field}.
\bjtitle{Comptes Rendus Geoscience}
\bvolume{340}(\bissue{12}),
\bfpage{791}--\blpage{800}
(\byear{2008})
\end{barticle}
\endbibitem

%%% 8
\bibitem{deighan2018discovery}
\begin{barticle}
\bauthor{\bsnm{Deighan}, \binits{J.}},
\bauthor{\bsnm{Jain}, \binits{S.}},
\bauthor{\bsnm{Chaffin}, \binits{M.}},
\bauthor{\bsnm{Fang}, \binits{X.}},
\bauthor{\bsnm{Halekas}, \binits{J.S.}},
\bauthor{\bsnm{Clarke}, \binits{J.T.}},
\bauthor{\bsnm{Schneider}, \binits{N.}},
\bauthor{\bsnm{Stewart}, \binits{A.}},
\bauthor{\bsnm{Chaufray}, \binits{J.-Y.}},
\bauthor{\bsnm{Evans}, \binits{J.}}, \betal:
\batitle{Discovery of a proton aurora at mars}.
\bjtitle{Nature Astronomy}
\bvolume{2}(\bissue{10}),
\bfpage{802}--\blpage{807}
(\byear{2018})
\end{barticle}
\endbibitem

%%% 9
\bibitem{bertaux2005discovery}
\begin{barticle}
\bauthor{\bsnm{Bertaux}, \binits{J.-L.}},
\bauthor{\bsnm{Leblanc}, \binits{F.}},
\bauthor{\bsnm{Witasse}, \binits{O.}},
\bauthor{\bsnm{Quemerais}, \binits{E.}},
\bauthor{\bsnm{Lilensten}, \binits{J.}},
\bauthor{\bsnm{Stern}, \binits{S.}},
\bauthor{\bsnm{Sandel}, \binits{B.}},
\bauthor{\bsnm{Korablev}, \binits{O.}}:
\batitle{Discovery of an aurora on mars}.
\bjtitle{Nature}
\bvolume{435}(\bissue{7043}),
\bfpage{790}--\blpage{794}
(\byear{2005})
\end{barticle}
\endbibitem

%%% 10
\bibitem{bertaux2006spicam}
\begin{botherref}
\oauthor{\bsnm{Bertaux}, \binits{J.-L.}},
\oauthor{\bsnm{Korablev}, \binits{O.}},
\oauthor{\bsnm{Perrier}, \binits{S.}},
\oauthor{\bsnm{Quemerais}, \binits{E.}},
\oauthor{\bsnm{Montmessin}, \binits{F.}},
\oauthor{\bsnm{Leblanc}, \binits{F.}},
\oauthor{\bsnm{Lebonnois}, \binits{S.}},
\oauthor{\bsnm{Rannou}, \binits{P.}},
\oauthor{\bsnm{Lef{\`e}vre}, \binits{F.}},
\oauthor{\bsnm{Forget}, \binits{F.}}, et al.:
Spicam on mars express: Observing modes and overview of uv spectrometer data
  and scientific results.
Journal of Geophysical Research: Planets
\textbf{111}(E10)
(2006)
\end{botherref}
\endbibitem

%%% 11
\bibitem{chicarro2004mars}
\begin{barticle}
\bauthor{\bsnm{Chicarro}, \binits{A.}},
\bauthor{\bsnm{Martin}, \binits{P.}},
\bauthor{\bsnm{Trautner}, \binits{R.}}:
\batitle{The mars express mission: an overview}.
\bjtitle{Mars Express: the scientific payload}
\bvolume{1240},
\bfpage{3}--\blpage{13}
(\byear{2004})
\end{barticle}
\endbibitem

%%% 12
\bibitem{brain2006origin}
\begin{botherref}
\oauthor{\bsnm{Brain}, \binits{D.}},
\oauthor{\bsnm{Halekas}, \binits{J.}},
\oauthor{\bsnm{Peticolas}, \binits{L.}},
\oauthor{\bsnm{Lin}, \binits{R.}},
\oauthor{\bsnm{Luhmann}, \binits{J.}},
\oauthor{\bsnm{Mitchell}, \binits{D.}},
\oauthor{\bsnm{Delory}, \binits{G.}},
\oauthor{\bsnm{Bougher}, \binits{S.}},
\oauthor{\bsnm{Acu{\~n}a}, \binits{M.}},
\oauthor{\bsnm{R{\`e}me}, \binits{H.}}:
On the origin of aurorae on mars.
Geophysical Research Letters
\textbf{33}(1)
(2006)
\end{botherref}
\endbibitem

%%% 13
\bibitem{jakosky2015mars}
\begin{barticle}
\bauthor{\bsnm{Jakosky}, \binits{B.M.}},
\bauthor{\bsnm{Lin}, \binits{R.P.}},
\bauthor{\bsnm{Grebowsky}, \binits{J.M.}},
\bauthor{\bsnm{Luhmann}, \binits{J.G.}},
\bauthor{\bsnm{Mitchell}, \binits{D.}},
\bauthor{\bsnm{Beutelschies}, \binits{G.}},
\bauthor{\bsnm{Priser}, \binits{T.}},
\bauthor{\bsnm{Acuna}, \binits{M.}},
\bauthor{\bsnm{Andersson}, \binits{L.}},
\bauthor{\bsnm{Baird}, \binits{D.}}, \betal:
\batitle{The mars atmosphere and volatile evolution (maven) mission}.
\bjtitle{Space Science Reviews}
\bvolume{195}(\bissue{1}),
\bfpage{3}--\blpage{48}
(\byear{2015})
\end{barticle}
\endbibitem

%%% 14
\bibitem{schneider2021discrete}
\begin{barticle}
\bauthor{\bsnm{Schneider}, \binits{N.}},
\bauthor{\bsnm{Milby}, \binits{Z.}},
\bauthor{\bsnm{Jain}, \binits{S.}},
\bauthor{\bsnm{G{\'e}rard}, \binits{J.-C.}},
\bauthor{\bsnm{Soret}, \binits{L.}},
\bauthor{\bsnm{Brain}, \binits{D.}},
\bauthor{\bsnm{Weber}, \binits{T.}},
\bauthor{\bsnm{Girazian}, \binits{Z.}},
\bauthor{\bsnm{McFadden}, \binits{J.}},
\bauthor{\bsnm{Deighan}, \binits{J.}}, \betal:
\batitle{Discrete aurora on mars: Insights into their distribution and activity
  from maven/iuvs observations}.
\bjtitle{Journal of Geophysical Research: Space Physics}
\bvolume{126}(\bissue{10}),
\bfpage{2021}--\blpage{029428}
(\byear{2021})
\end{barticle}
\endbibitem

%%% 15
\bibitem{xu2022empirically}
\begin{barticle}
\bauthor{\bsnm{Xu}, \binits{S.}},
\bauthor{\bsnm{Mitchell}, \binits{D.L.}},
\bauthor{\bsnm{McFadden}, \binits{J.P.}},
\bauthor{\bsnm{Schneider}, \binits{N.M.}},
\bauthor{\bsnm{Milby}, \binits{Z.}},
\bauthor{\bsnm{Jain}, \binits{S.}},
\bauthor{\bsnm{Weber}, \binits{T.}},
\bauthor{\bsnm{Brain}, \binits{D.A.}},
\bauthor{\bsnm{DiBraccio}, \binits{G.A.}},
\bauthor{\bsnm{Halekas}, \binits{J.}}, \betal:
\batitle{Empirically determined auroral electron events at mars—maven
  observations}.
\bjtitle{Geophysical Research Letters}
\bvolume{49}(\bissue{6}),
\bfpage{2022}--\blpage{097757}
(\byear{2022})
\end{barticle}
\endbibitem

%%% 16
\bibitem{girazian2022discrete}
\begin{barticle}
\bauthor{\bsnm{Girazian}, \binits{Z.}},
\bauthor{\bsnm{Schneider}, \binits{N.M.}},
\bauthor{\bsnm{Milby}, \binits{Z.}},
\bauthor{\bsnm{Fang}, \binits{X.}},
\bauthor{\bsnm{Halekas}, \binits{J.}},
\bauthor{\bsnm{Weber}, \binits{T.}},
\bauthor{\bsnm{Jain}, \binits{S.}},
\bauthor{\bsnm{G{\'e}rard}, \binits{J.-C.}},
\bauthor{\bsnm{Soret}, \binits{L.}},
\bauthor{\bsnm{Deighan}, \binits{J.}}, \betal:
\batitle{Discrete aurora at mars: Dependence on upstream solar wind
  conditions}.
\bjtitle{Journal of Geophysical Research: Space Physics}
\bvolume{127}(\bissue{4}),
\bfpage{2021}--\blpage{030238}
(\byear{2022})
\end{barticle}
\endbibitem

%%% 17
\bibitem{schneider2015discovery}
\begin{barticle}
\bauthor{\bsnm{Schneider}, \binits{N.M.}},
\bauthor{\bsnm{Deighan}, \binits{J.I.}},
\bauthor{\bsnm{Jain}, \binits{S.K.}},
\bauthor{\bsnm{Stiepen}, \binits{A.}},
\bauthor{\bsnm{Stewart}, \binits{A.I.F.}},
\bauthor{\bsnm{Larson}, \binits{D.}},
\bauthor{\bsnm{Mitchell}, \binits{D.L.}},
\bauthor{\bsnm{Mazelle}, \binits{C.}},
\bauthor{\bsnm{Lee}, \binits{C.O.}},
\bauthor{\bsnm{Lillis}, \binits{R.J.}}, \betal:
\batitle{Discovery of diffuse aurora on mars}.
\bjtitle{Science}
\bvolume{350}(\bissue{6261}),
\bfpage{0313}
(\byear{2015})
\end{barticle}
\endbibitem

%%% 18
\bibitem{mcclintock2015imaging}
\begin{barticle}
\bauthor{\bsnm{McClintock}, \binits{W.E.}},
\bauthor{\bsnm{Schneider}, \binits{N.M.}},
\bauthor{\bsnm{Holsclaw}, \binits{G.M.}},
\bauthor{\bsnm{Clarke}, \binits{J.T.}},
\bauthor{\bsnm{Hoskins}, \binits{A.C.}},
\bauthor{\bsnm{Stewart}, \binits{I.}},
\bauthor{\bsnm{Montmessin}, \binits{F.}},
\bauthor{\bsnm{Yelle}, \binits{R.V.}},
\bauthor{\bsnm{Deighan}, \binits{J.}}:
\batitle{The imaging ultraviolet spectrograph (iuvs) for the maven mission}.
\bjtitle{Space Science Reviews}
\bvolume{195}(\bissue{1}),
\bfpage{75}--\blpage{124}
(\byear{2015})
\end{barticle}
\endbibitem

%%% 19
\bibitem{ritter2018proton}
\begin{barticle}
\bauthor{\bsnm{Ritter}, \binits{B.}},
\bauthor{\bsnm{Gérard}, \binits{J.-C.}},
\bauthor{\bsnm{Hubert}, \binits{B.}},
\bauthor{\bsnm{Rodriguez}, \binits{L.}},
\bauthor{\bsnm{Montmessin}, \binits{F.}}:
\batitle{Observations of the proton aurora on mars with spicam on board mars
  express}.
\bjtitle{Geophysical Research Letters}
\bvolume{45}(\bissue{2}),
\bfpage{612}--\blpage{619}
(\byear{2018}).
\doiurl{10.1002/2017GL076235}
\end{barticle}
\endbibitem

%%% 20
\bibitem{holsclaw2021emirates}
\begin{barticle}
\bauthor{\bsnm{Holsclaw}, \binits{G.M.}},
\bauthor{\bsnm{Deighan}, \binits{J.}},
\bauthor{\bsnm{Almatroushi}, \binits{H.}},
\bauthor{\bsnm{Chaffin}, \binits{M.}},
\bauthor{\bsnm{Correira}, \binits{J.}},
\bauthor{\bsnm{Evans}, \binits{J.S.}},
\bauthor{\bsnm{Fillingim}, \binits{M.}},
\bauthor{\bsnm{Hoskins}, \binits{A.}},
\bauthor{\bsnm{Jain}, \binits{S.K.}},
\bauthor{\bsnm{Lillis}, \binits{R.}}, \betal:
\batitle{The emirates mars ultraviolet spectrometer (emus) for the emm
  mission}.
\bjtitle{Space Science Reviews}
\bvolume{217}(\bissue{8}),
\bfpage{1}--\blpage{49}
(\byear{2021})
\end{barticle}
\endbibitem

%%% 21
\bibitem{chaffin2022patchy}
\begin{barticle}
\bauthor{\bsnm{Chaffin}, \binits{M.S.}},
\bauthor{\bsnm{Fowler}, \binits{C.M.}},
\bauthor{\bsnm{Deighan}, \binits{J.}},
\bauthor{\bsnm{Jain}, \binits{S.}},
\bauthor{\bsnm{Holsclaw}, \binits{G.}},
\bauthor{\bsnm{Hughes}, \binits{A.}},
\bauthor{\bsnm{Ramstad}, \binits{R.}},
\bauthor{\bsnm{Dong}, \binits{Y.}},
\bauthor{\bsnm{Brain}, \binits{D.}},
\bauthor{\bsnm{AlMazmi}, \binits{H.}}, \betal:
\batitle{Patchy proton aurora at mars: A global view of solar wind
  precipitation across the martian dayside from emm/emus}.
\bjtitle{Geophysical Research Letters}
\bvolume{49}(\bissue{17}),
\bfpage{2022}--\blpage{099881}
(\byear{2022})
\end{barticle}
\endbibitem

%%% 22
\bibitem{halekas2015wind}
\begin{barticle}
\bauthor{\bsnm{Halekas}, \binits{J.S.}},
\bauthor{\bsnm{Lillis}, \binits{R.J.}},
\bauthor{\bsnm{Mitchell}, \binits{D.L.}},
\bauthor{\bsnm{Cravens}, \binits{T.E.}},
\bauthor{\bsnm{Mazelle}, \binits{C.}},
\bauthor{\bsnm{Connerney}, \binits{J.E.P.}},
\bauthor{\bsnm{Espley}, \binits{J.R.}},
\bauthor{\bsnm{Mahaffy}, \binits{P.R.}},
\bauthor{\bsnm{Benna}, \binits{M.}},
\bauthor{\bsnm{Jakosky}, \binits{B.M.}},
\bauthor{\bsnm{Luhmann}, \binits{J.G.}},
\bauthor{\bsnm{McFadden}, \binits{J.P.}},
\bauthor{\bsnm{Larson}, \binits{D.E.}},
\bauthor{\bsnm{Harada}, \binits{Y.}},
\bauthor{\bsnm{Ruhunusiri}, \binits{S.}}:
\batitle{Maven observations of solar wind hydrogen deposition in the atmosphere
  of mars}.
\bjtitle{Geophysical Research Letters}
\bvolume{42}(\bissue{21}),
\bfpage{8901}--\blpage{8909}
(\byear{2015}).
\doiurl{10.1002/2015GL064693}
\end{barticle}
\endbibitem

%%% 23
\bibitem{hughes2019proton}
\begin{barticle}
\bauthor{\bsnm{Hughes}, \binits{A.}},
\bauthor{\bsnm{Chaffin}, \binits{M.}},
\bauthor{\bsnm{Mierkiewicz}, \binits{E.}},
\bauthor{\bsnm{Deighan}, \binits{J.}},
\bauthor{\bsnm{Jain}, \binits{S.}},
\bauthor{\bsnm{Schneider}, \binits{N.}},
\bauthor{\bsnm{Mayyasi}, \binits{M.}},
\bauthor{\bsnm{Jakosky}, \binits{B.}}:
\batitle{Proton aurora on mars: A dayside phenomenon pervasive in southern
  summer}.
\bjtitle{Journal of Geophysical Research: Space Physics}
\bvolume{124}(\bissue{12}),
\bfpage{10533}--\blpage{10548}
(\byear{2019}).
\doiurl{10.1029/2019JA027140}
\end{barticle}
\endbibitem

%%% 24
\bibitem{Halekas2017variability}
\begin{barticle}
\bauthor{\bsnm{Halekas}, \binits{J.S.}},
\bauthor{\bsnm{Ruhunusiri}, \binits{S.}},
\bauthor{\bsnm{Harada}, \binits{Y.}},
\bauthor{\bsnm{Collinson}, \binits{G.}},
\bauthor{\bsnm{Mitchell}, \binits{D.L.}},
\bauthor{\bsnm{Mazelle}, \binits{C.}},
\bauthor{\bsnm{McFadden}, \binits{J.P.}},
\bauthor{\bsnm{Connerney}, \binits{J.E.P.}},
\bauthor{\bsnm{Espley}, \binits{J.R.}},
\bauthor{\bsnm{Eparvier}, \binits{F.}},
\bauthor{\bsnm{Luhmann}, \binits{J.G.}},
\bauthor{\bsnm{Jakosky}, \binits{B.M.}}:
\batitle{Structure, dynamics, and seasonal variability of the mars-solar wind
  interaction: Maven solar wind ion analyzer in-flight performance and science
  results}.
\bjtitle{Journal of Geophysical Research: Space Physics}
\bvolume{122}(\bissue{1}),
\bfpage{547}--\blpage{578}
(\byear{2017}).
\doiurl{10.1002/2016JA023167}
\end{barticle}
\endbibitem

%%% 25
\bibitem{GERARD2019mag}
\begin{barticle}
\bauthor{\bsnm{Gérard}, \binits{J.C.}},
\bauthor{\bsnm{Hubert}, \binits{B.}},
\bauthor{\bsnm{Ritter}, \binits{B.}},
\bauthor{\bsnm{Shematovich}, \binits{V.I.}},
\bauthor{\bsnm{Bisikalo}, \binits{D.V.}}:
\batitle{Lyman-alpha emission in the martian proton aurora: Line profile and
  role of horizontal induced magnetic field}.
\bjtitle{Icarus}
\bvolume{321},
\bfpage{266}--\blpage{271}
(\byear{2019}).
\doiurl{10.1016/j.icarus.2018.11.013}
\end{barticle}
\endbibitem

%%% 26
\bibitem{Hughes2021AGU}
\begin{bchapter}
\bauthor{\bsnm{{Hughes}}, \binits{A.}},
\bauthor{\bsnm{{Chaffin}}, \binits{M.}},
\bauthor{\bsnm{{Mierkiewicz}}, \binits{E.}},
\bauthor{\bsnm{{DiBraccio}}, \binits{G.}},
\bauthor{\bsnm{{Schneider}}, \binits{N.}},
\bauthor{\bsnm{{Deighan}}, \binits{J.}},
\bauthor{\bsnm{{Jain}}, \binits{S.}},
\bauthor{\bsnm{{Jakosky}}, \binits{B.}}:
\bctitle{{Evaluating the Influence of the Martian Magnetic Field Environment on
  Proton Aurora Activity}}.
In: \bbtitle{AGU Fall Meeting Abstracts},
vol. \bseriesno{2021},
pp. \bfpage{44}--\blpage{02}
(\byear{2021})
\end{bchapter}
\endbibitem

%%% 27
\bibitem{amiri2022emirates}
\begin{barticle}
\bauthor{\bsnm{Amiri}, \binits{H.}},
\bauthor{\bsnm{Brain}, \binits{D.}},
\bauthor{\bsnm{Sharaf}, \binits{O.}},
\bauthor{\bsnm{Withnell}, \binits{P.}},
\bauthor{\bsnm{McGrath}, \binits{M.}},
\bauthor{\bsnm{Alloghani}, \binits{M.}},
\bauthor{\bsnm{Al~Awadhi}, \binits{M.}},
\bauthor{\bsnm{Al~Dhafri}, \binits{S.}},
\bauthor{\bsnm{Al~Hamadi}, \binits{O.}},
\bauthor{\bsnm{Al~Matroushi}, \binits{H.}}, \betal:
\batitle{The emirates mars mission}.
\bjtitle{Space Science Reviews}
\bvolume{218}(\bissue{1}),
\bfpage{1}--\blpage{46}
(\byear{2022})
\end{barticle}
\endbibitem

%%% 28
\bibitem{almatroushi2021emirates}
\begin{barticle}
\bauthor{\bsnm{Almatroushi}, \binits{H.}},
\bauthor{\bsnm{AlMazmi}, \binits{H.}},
\bauthor{\bsnm{AlMheiri}, \binits{N.}},
\bauthor{\bsnm{AlShamsi}, \binits{M.}},
\bauthor{\bsnm{AlTunaiji}, \binits{E.}},
\bauthor{\bsnm{Badri}, \binits{K.}},
\bauthor{\bsnm{Lillis}, \binits{R.J.}},
\bauthor{\bsnm{Lootah}, \binits{F.}},
\bauthor{\bsnm{Yousuf}, \binits{M.}},
\bauthor{\bsnm{Amiri}, \binits{S.}}, \betal:
\batitle{Emirates mars mission characterization of mars atmosphere dynamics and
  processes}.
\bjtitle{Space Science Reviews}
\bvolume{217}(\bissue{8}),
\bfpage{1}--\blpage{31}
(\byear{2021})
\end{barticle}
\endbibitem

%%% 29
\bibitem{2022AoM}
\begin{bbook}
\bauthor{\bsnm{Atri}, \binits{D.}},
\bauthor{\bsnm{Alhantoobi}, \binits{A.}},
\bauthor{\bsnm{Fialova}, \binits{K.}},
\bauthor{\bsnm{Singh}, \binits{S.P.}},
\bauthor{\bsnm{Dhuri}, \binits{D.B.}},
\bauthor{\bsnm{Abdelmoneim}, \binits{N.}}:
\bbtitle{Atlas of Mars: A Photographic Atlas, Prepared with Observations from
  the Emirates Mars Mission, or ``Hope" (al-amal)}.
\bpublisher{New York University Abu Dhabi Center for Space Science},
\blocation{Abu Dhabi, 100 pp}
(\byear{2022})
\end{bbook}
\endbibitem

%%% 30
\bibitem{atri2022diurnal}
\begin{botherref}
\oauthor{\bsnm{Atri}, \binits{D.}},
\oauthor{\bsnm{Abdelmoneim}, \binits{N.}},
\oauthor{\bsnm{Dhuri}, \binits{D.B.}},
\oauthor{\bsnm{Simoni}, \binits{M.}}:
Diurnal variation of the surface temperature of mars with the emirates mars
  mission: A comparison with curiosity and perseverance rover measurements.
arXiv preprint arXiv:2204.12850
(2022)
\end{botherref}
\endbibitem

%%% 31
\bibitem{lillisfirst}
\begin{botherref}
\oauthor{\bsnm{Lillis}, \binits{R.J.}},
\oauthor{\bsnm{Deighan}, \binits{J.}},
\oauthor{\bsnm{Brain}, \binits{D.}},
\oauthor{\bsnm{Fillingim}, \binits{M.}},
\oauthor{\bsnm{Jain}, \binits{S.}},
\oauthor{\bsnm{Chaffin}, \binits{M.}},
\oauthor{\bsnm{Holsclaw2}, \binits{G.} \bsuffix{Krishnaprasad~Chirakkil}},
\oauthor{\bsnm{Al~Matroushi}, \binits{H.}},
\oauthor{\bsnm{Lootah}, \binits{F.}},
\oauthor{\bsnm{Al~Mazmi}, \binits{H.}}, et al.:
First synoptic images of fuv discrete aurora and discovery of sinuous aurora at
  mars by emm emus.
Geophysical Research Letters,
2022--099820
\end{botherref}
\endbibitem

%%% 32
\bibitem{Gao2021crustal}
\begin{barticle}
\bauthor{\bsnm{Gao}, \binits{J.W.}},
\bauthor{\bsnm{Rong}, \binits{Z.J.}},
\bauthor{\bsnm{Klinger}, \binits{L.}},
\bauthor{\bsnm{Li}, \binits{X.Z.}},
\bauthor{\bsnm{Liu}, \binits{D.}},
\bauthor{\bsnm{Wei}, \binits{Y.}}:
\batitle{A spherical harmonic martian crustal magnetic field model combining
  data sets of maven and mgs}.
\bjtitle{Earth and Space Science}
\bvolume{8}(\bissue{10}),
\bfpage{2021}--\blpage{001860}
(\byear{2021}).
\doiurl{10.1029/2021EA001860}.
\bcomment{e2021EA001860 2021EA001860}
\end{barticle}
\endbibitem

%%% 33
\bibitem{crismani2019ionosphere}
\begin{barticle}
\bauthor{\bsnm{Crismani}, \binits{M.M.J.}},
\bauthor{\bsnm{Deighan}, \binits{J.}},
\bauthor{\bsnm{Schneider}, \binits{N.M.}},
\bauthor{\bsnm{Plane}, \binits{J.M.C.}},
\bauthor{\bsnm{Withers}, \binits{P.}},
\bauthor{\bsnm{Halekas}, \binits{J.}},
\bauthor{\bsnm{Chaffin}, \binits{M.}},
\bauthor{\bsnm{Jain}, \binits{S.}}:
\batitle{Localized ionization hypothesis for transient ionospheric layers}.
\bjtitle{Journal of Geophysical Research: Space Physics}
\bvolume{124}(\bissue{6}),
\bfpage{4870}--\blpage{4880}
(\byear{2019}).
\doiurl{10.1029/2018JA026251}
\end{barticle}
\endbibitem

\end{thebibliography}


\newpage

 \appendix
 \onecolumn
 \counterwithin{figure}{section}
 
 %\section
 \vspace{2.0cm}
 \section{\bf {\Large{Appendix: Observations of discrete auroral events}}}
  \makeatletter
 \def\@captype{figure}
 \makeatother
%\begin{figure*}
\centering
 	\includegraphics[width=12cm]{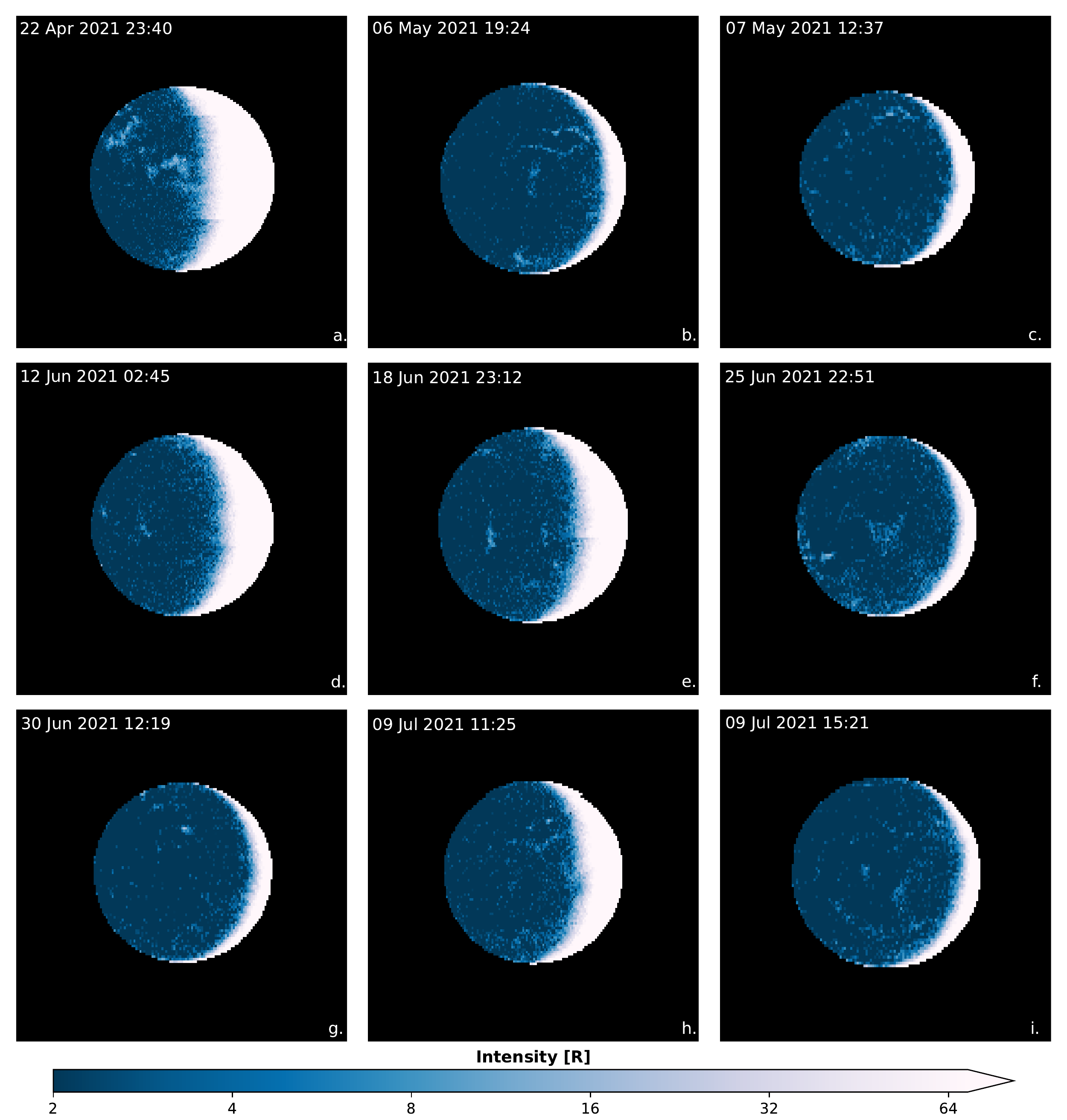}
     \caption{Observations of auroras by the EMUS instrument in the 130.4 nm oxygen band (part 1).}
     \label{fig:auroras_observations_appendix1}
%\end{figure*}

 \newpage
  \makeatletter
 \def\@captype{figure}
 \makeatother
% \begin{figure*}
\centering
 	\includegraphics[width=12cm]{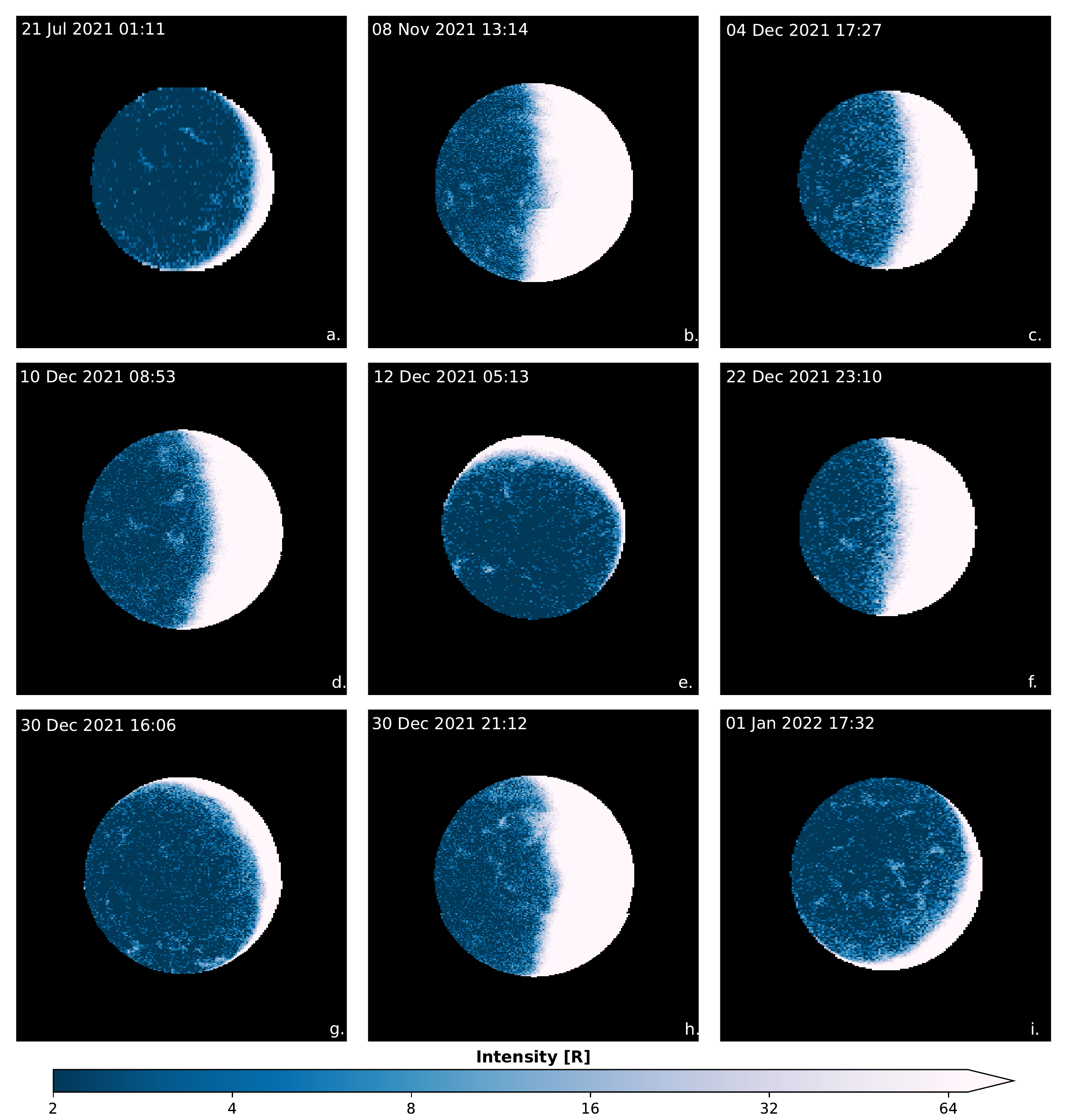}
     \caption{Observations of auroras by the EMUS instrument in the 130.4 nm oxygen band (part 2).}
     \label{fig:auroras_observations_appendix2}
% \end{figure*}

\newpage
 \makeatletter
\def\@captype{figure}
\makeatother
% \begin{figure*}
\centering
 	\includegraphics[width=8cm]{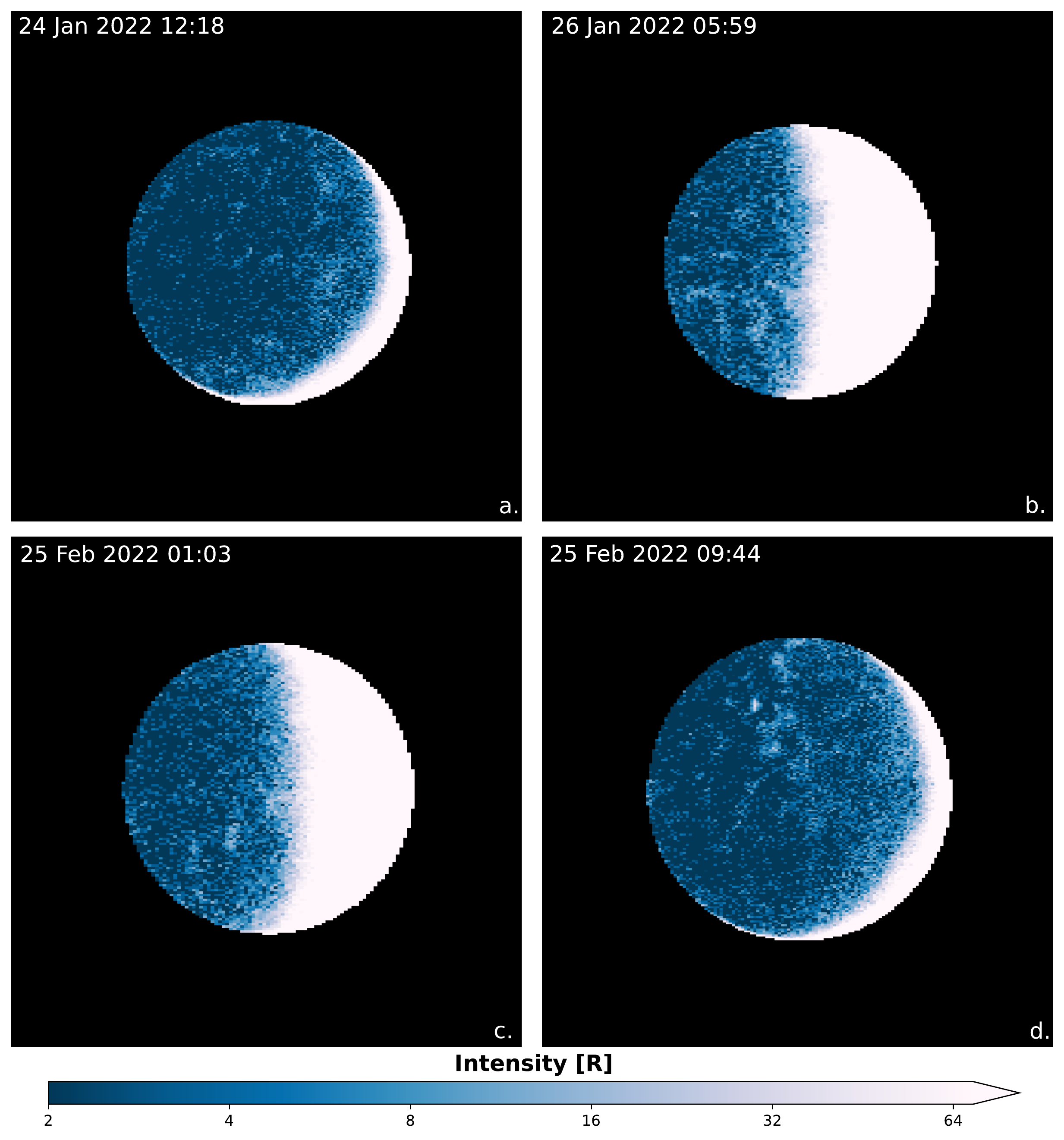}
     \caption{Observations of auroras by the EMUS instrument in the 130.4 nm oxygen band (part 3).}
     \label{fig:auroras_observations_appendix3}
% \end{figure*}
\end{document}